\documentstyle[12pt]{article}

\global\arraycolsep=1pt
\def\be{\begin{equation}}
\def\ee{\end{equation}}
\def\bea{\begin{eqnarray}}
\def\eea{\end{eqnarray}}
\def\al{\alpha}
\def\tal{{\tilde \alpha}}
\def\tbe{{\tilde \beta}}
\def\ep{\epsilon}
\def\ve{\varepsilon}
\def\sg{\sigma}
\def\zt{\zeta}
\def\tht{\theta}
\def\gm{\gamma}
\def\Gm{\Gamma}
\def\lm{\lambda}
\def\dlt{\Delta}
\def\t0{{\tilde 0}}
\def\hf{{1 \over 2}}
\def\hsf{{1 \over {\sqrt 2}}}
\newfont{\eufm}{eufm10 scaled \magstep1}
\begin{document}
\begin{flushright}
OU-HET 322 \\ {\bf hep-th/9909007} 
\end{flushright}
\vskip 1.5 cm

\begin{Large}
\centerline{{\bf BPS States in 10+2 Dimensions}}
\end{Large}
\vskip 2.0cm

\centerline{{\bf Tatsuya Ueno}}
\vskip 1.0cm

\centerline{Department of Physics, Graduate School of 
Science,}
\vskip 5pt
\centerline{Osaka University}
\vskip 5pt
\centerline{Toyonaka, Osaka 560-0043, Japan.}
\vskip 5pt
\centerline{\quad{\sl ueno@funpth.phys.sci.osaka-u.ac.jp}} 
\vskip 3.0cm

\begin{center}
{\bf ABSTRACT}
\end{center}
We discuss a (10+2)D $N=(1,1)$ superalgebra and its projections to 
M-theory, type IIA and IIB algebras.
{}From the complete classification of a second-rank central term valued 
in the $so(10,2)$ algebra, we find all possible BPS states 
coming from this term.
We show that, among them, there are two types of $1/2$-susy BPS 
configurations; one corresponds to a super (2+2)-brane while another one 
arises from a nilpotent element in $so(10,2)$.
\thispagestyle{empty}
\newpage

\section{Introduction}%
It is intriguing to consider the possibility that a Theory of Everything
predicts not only the dimensionality of spacetime but also its signature. 
In a paper \cite{bd}, with this spirit in mind, Blencowe and Duff 
investigated possible supersymmetric branes moving in spacetime 
without {\it a priori} assuming definite signatures of both the brane 
worldvolume and the spacetime.
Requiring local $\kappa$-symmetry on the worldvolume and 
spacetime supersymmetry, they found, adding to well known string 
and membrane configurations, two extra octonionic type branes;
a (1+2)-brane in (9+2)D and a (2+2)-brane in (10+2)D.
In \cite{hk}, the (1+2)-brane was shown to be realized as a solution 
of a (9+2)D theory obtained from the usual (10+1)D supergravity by 
spacelike and timelike T-duality transformations.
The latter (2+2)-brane can possibly exist only when we consider
an $N=(1,0)$ non-Poincar\'e spacetime supersymmetry,
which is the maximal symmetry to have 32 real supercharges.
\par

In recent development of string unification, the connection of  
various supersymmetric theories in higher dimensions has been 
discussed.
It is well known that the conjectured M-theory, when compactified on 
$S^1$ and $S^1/Z_2$, gives type IIA and $E_8 \times E_8$ heterotic 
string theories, respectively \cite{pt1}.
M-theory also leads to type IIB, $SO(32)$ heterotic and type I 
strings less directly upon compactification to nine dimensions. 
Other higher dimensional structures also have been explored to obtain
a unified picture of those strings and to explain various duality 
relations among them, in which the notion of extra time dimensions 
has emerged as hidden dimensions of the higher dimensional unification 
theories \cite{cv},\cite{ib}.
A notable example is F-theory \cite{cv} which has been proposed as 
a (10+2)D structure to give a geometrical explanation of the self 
S-duality of the type IIB string.
In \cite{rss}, superalgebras in dimensions beyond eleven have been
studied in the context of the unification of M-theory, type IIA and 
IIB algebras. 
With the restriction of the number of real supercharges to be 64 or 
less, they found two possible distinct superalgebras; $N=(2,0)$ algebra 
in (10+2)D and $N=(1,0)$ algebra in (11+3)D, where the latter is 
reduced to various algebras in lower dimensions; $N=1$ algebras in
(11+2)D, (10+3)D and (11+1)D, $N=2$ algebra in (10+1)D and $N=(1,1)$ 
algebra in (10+2)D.
\par

A fundamental question about physics with two or more time dimensions
is what kind of local theory can exist in such a spacetime.
There have been many suggestions concerning it \cite{hp}-\cite{sh2}.
We should be careful to discuss the dynamical aspect of 
such a theory, since the concept of dynamics and energy used in 
ordinary theories with one time should be modified in the present case.
Alternatively it would be a promising way to start with the analysis
of the symmetrical aspect of the theory.
Indeed, the structure of superalgebra is sensitive to the dimensionality 
and the signature of spacetime and would teach us about the possibility
of extra time directions.
It also seems possible to see what types of fundamental objects
exist in the superalgebra through the investigation of central 
terms appearing in anti-commutators among supercharges.
\par

In this paper, we discuss $N=(1,1)$ and $(1,0)$ superalgebras in 
(10+2)D, assuming that the algebras are relevant to some local theory 
in (10+2)D.
As described in Section 2, the $N=(1,1)$ algebra is composed of the 
$SO(10,2)$ generators $M_{AB}$, (pseudo)-Majorana spinor supercharges 
$Q^{\al}$ and tensorial central terms $Z^{(k)}_{A_1 \cdots A_k}$ 
$(k=2,3,6,7,10,11)$ \cite{ib},\cite{hp}.
Since the algebra does not contain any vector generator, it is {\it not}
the Poincar\'e type superalgebra, so is rather unfamiliar to us. 
However, we see that its Weyl projected $N=(1,0)$ algebra is related to 
the (10+1)D M-theory algebra (M-algebra) and the (9+1)D type IIA algebra 
by dimensional reduction.
It is also shown that another projection of the (10+2)D $N=(1,1)$
algebra, together with dimensional reduction to (9+1)D, leads to the 
type IIB algebra.
\par

A primary interest on the (10+2)D algebras is how many and what types of 
BPS configurations possibly exist in them.
The above mentioned connection of the (10+2)D $N=(1,1)$ algebra to M-, 
type IIA and IIB algebras may suggest a (10+2)D origin of various BPS 
states in M-theory and all known superstring theories.
In this paper, we concentrate on the second-rank tensorial central term 
$Z^{(2)}_{AB}$ and investigate possible BPS states arising from this 
term.
As a first step, we need to simplify the form of the central term.
$Z^{(2)}$ belongs to the $so(10,2)$ algebra and hence transforms
under the adjoint action of the $SO(10,2)$ group. 
Unlike the cases of complex or real compact Lie algebras, $so(10,2)$ 
elements are not reduced to a unique form under the $SO(10,2)$ action
and hence the classification of $so(10,2)$ becomes complicated.
With the use of results in \cite{ms},\cite{lr},\cite{bc}, we 
classify conjugacy classes of $so(10,2)$ completely and construct 
representatives for the classes, which contain at most 6 parameters.
The classification enables us to pick up all possible BPS states 
characterized by some fraction of surviving supersymmetry, as
shown in Section 3.
It is also demonstrated how the (10+2)D BPS states reduce to 
those in M-theory and type IIA and IIB theories.
Finally several concluding remarks are given in Section 4.
\par

The (10+2)D flat metric is $\eta_{AB} = (-,-,+,\cdots,+)$.
$A, B, \cdots =\t0,0,1,\cdots,10$ denotes the (10+2)D indices, while 
$a,b,\cdots=0,1,\cdots,10$ and $\mu,\nu,\cdots=0,1,\cdots,9$ are 
(10+1)D and (9+1)D indices, respectively.
\par

\newpage
\section{$N=(1,1)$ superalgebra in (10+2)D}%
We consider a graded generalization of the $SO(10,2)$ Lorentz 
algebra with generators $M_{AB}$ in terms of spinor supersymmetry 
generators $Q^{\al}$, 
\bea
&&[M_{AB}, M_{CD}] = 
M_{AC}\eta_{BD} + M_{BD}\eta_{AC} -
M_{AD}\eta_{BC} - M_{BC}\eta_{AD} \ , \label{mm}\\
&&[M_{AB}, Q^{\al}] = - \hf (\Gm_{AB})^{\al}_{\ \beta} \,
Q^{\beta} \ .
\label{mq}
\eea
With these commutation relations, we have to define anti-commutators among 
the $Q^{\al}$'s, which depends on the dimensionality and the signature of 
spacetime.
\par

\subsection{Spinors in $D=(S,T)$}%
In a general $D=(S,T)$ spacetime, with gamma matrices $\Gm_A$,
\be
\{\Gm_A, \Gm_B\} = 2 \, \eta_{AB} I_D \ , 
\ee
it can be shown that there exists a matrix $B$ satisfying 
\be
\Gm_A = \kappa B^{-1} \Gm_A^{\ast} B \ ,  \qquad
B^{\ast} B = \ep I_D \ ,
\ee
where parameters $\kappa$ and $\ep$ are specified by the value
of $S-T$ as follows \cite{kt}.
\begin{center} Table 1. \\
\begin{tabular}[b]{|c|c|c|c|c|} \hline
$S-T$ \ mod 8 & 0, 1, 2 & 0, 6, 7 & 4, 5, 6 & 2, 3, 4 \\ \hline
$\ep$        & $+1$ & $+1$ & $-1$ & $-1$ \\
$\kappa$   & $+1$ & $-1$  & $-1$ & $+1$ \\ \hline
\end{tabular}
\end{center}
With the product of all time component gamma matrices 
$A = \Gm^{0_1}\cdots\Gm^{0_T}$, 
the charge conjugation matrix is defined as $C = {\tilde B} A$, which satisfies
\be
{\tilde \Gm}_A = (-1)^T \kappa \, C \Gm_A C^{-1} \ , \quad
{\tilde C} = \ep \, \kappa^T (-1)^{T(T+1)/2} C \ ,
\label{cm}
\ee
where the tilde denotes transpose.
In the $\ep = +1$ case, we can impose the Majorana $(\kappa = -1)$
or the pseudo-Majorana $(\kappa=+1)$ condition as
\be
Q^* = B Q \quad \Longleftrightarrow \quad 
{\bar Q} \equiv Q^{\dagger} A = {\tilde Q} C \ .
\label{mjc}
\ee
In the even dimension $D$, we can define a definite chirality for 
spinors by the chirality matrix
\be
\Gm \equiv (-1)^{(S-T)/4} \Gm^{0_1}\cdots\Gm^{0_T}\Gm^1 \cdots \Gm^S \ , \quad
\Gm^2 = I_D \ , \quad \{\Gm , \Gm^A \} = 0 \ .
\label{ch}
\ee
{}From (\ref{cm}) and (\ref{ch}) , we obtain the condition 
$\Gm = B^{-1} \Gm^{\ast} B$, if $S-T=0$ mod 4, showing that 
it is possible to impose the (pseudo)-Majorana-Weyl condition for
spinors in 10+2 dimensions.

\subsection{Anti-commutators}%
The general form of anti-commutation relations among the 
(pseudo-)Majorana spinors $Q^{\al}$ becomes
\be
\{Q^{\al}, Q^{\beta}\} = 
\sum_k {1 \over k!} (\Gm^{A_1 \cdots A_k}C^{-1})^{\al\beta}
Z^{(k)}_{A_1 \cdots A_k} \ .
\label{gqq}
\ee 
Since the left-hand-side (LHS) of (\ref{gqq})  is a real symmetric matrix, 
we have to choose $Z^{(k)}$ which make the right-hand-side (RHS) of 
(\ref{gqq}) symmetric.
In 10+2 dimensions, using (\ref{cm}), we can show that
\bea
\Gm^{A_1 \cdots A_k}C^{-1} {\rm \ is \ symmetric} &&
{\rm for} \ k=1,2 \ {\rm mod} \ 4 {\rm \ for \ the} \ C_-  \ {\rm case}, 
\nonumber \\
&&{\rm for} \ k=2,3 \ {\rm mod} \ 4 {\rm \ for \ the} \ C_+ \ {\rm case}.
\nonumber 
\eea
\par

For the $C_-$ case, the algebra consists of the set of generators 
$\{M_{AB}, Q^{\al}, Z^{(k)}\}$, $(k=1,2,5,6,9,10)$, which becomes a 
supersymmetric extension of the Poincar\'e algebra if the vector 
$Z^{(1)}_A$ is identified with the translation generator $P_A$.
This, however, causes a problem in the construction of the worldvolume 
theory \cite{hp}.
Since the matrix $(\Gm_{ABC}C_-^{-1})^{\al \beta}$ is anti-symmetric 
with respect to $(\al, \beta)$, the Wess-Zumino term on the (2+2)D 
worldvolume is identically zero, which means that the $\kappa$-symmetry 
can not be defined in this case.
Thus the degrees of freedom of bosons and fermions on the worldvolume 
do not match up.
\par

In the following, we consider the $C_+$ case which leads to a 
non-Poincar\'e superalgebra defined with (\ref{mm}), (\ref{mq}) and 
\be
\{Q^{\al}, Q^{\beta}\} = 
\, \hf \, (\Gm^{AB}C_+^{-1})^{\al\beta}Z^{(2)}_{AB} \, +
\sum_{k=3,6,7,10,11} {1 \over k!} \, 
(\Gm^{A_1 \cdots A_k}C_+^{-1})^{\al\beta} Z^{(k)}_{A_1 \cdots A_k} \ ,
\label{qq}
\ee 
where spinor indices $\al$, $\beta$ run from 1 to $2^{[12/2]}=64$.
The two-rank tensorial central term $Z^{(2)}_{AB}$ is supposed to be 
independent from the Lorentz generators $M_{AB}$. 
We note that $Z^{(2)}_{AB}$ has a constraint arising from the positivity 
condition that all eigenvalues $\lm_i$ of the matrix $\{Q,Q\}$ of (\ref{qq}) 
are non-negative.
Henceforth, for simplicity, we take $C_+ = \Gm^{\t0}\Gm^0$ $(B_+ = I)$ 
and the spinor $Q^{\al}$ to be real.
Then it can be easily shown that the trace of all terms in the RHS of 
(\ref{qq}) vanishes except for the term $(\Gm^{\t0 0}C_+^{-1}) Z_{\t0 0}$,
\be
{\rm Tr}\{Q^{\al},Q^{\beta}\} = \sum_{i=1}^{64} \lm_i = 
64 \, Z^{(2)}_{\t0 0} \geq 0 \ .
\label{pos}
\ee
\par

In the following, we demonstrate two projections of the (10+2)D spin space 
using the explicit representation of gamma matrices,
\be
\Gm^{\t0} = i \sg_2 \otimes I_{32} \ , \qquad
\Gm^a = \sg_1 \otimes \gm^a \ ,
\label{gm1}
\ee
where $\gm^a$ are (10+1)D gamma matrices, in which
$\gm^{\mu}$ equal to (9+1)D gamma matrices and 
$\gm^{10}= \gm^0 \cdots \gm^9$,
\be
\gm^0 = i \sg_2 \otimes I_{16} \ , \qquad
\gm^k = \sg_1 \otimes \gm^k_{(9)} \ , \ \ \ (k=1,\cdots,9)
\label{gm2}
\ee
where $\gm^k_{(9)}$ are 9D real symmetric gamma matrices and $\sg_i$ are 
Pauli matrices.
\par

\subsection{Weyl projection}%
We impose the Weyl condition to $Q^{\al}$ by the projection
operator $P_{\pm} \equiv \hf (I_{64} \pm \Gm)$, where 
$\Gm = \Gm^{\t0}\Gm^0\cdots \Gm^{10}$ in (\ref{ch}).
In this projection, $Z^{(k)}$ with $k=3,7,11$
drop out from (\ref{qq}) and anti-commutation relations
for Majorana-Weyl spinors 
$Q_{\pm}^{\al} = (P_{\pm})^{\al}_{\ \beta}Q^{\beta}$ become
\be
\{Q_{\pm}^{\al}, Q_{\pm}^{\beta}\} = 
\hf (P_{\pm}\Gm^{AB}C^{-1})^{\al\beta}T^{(\mp)}_{AB}
+ {1 \over 6!}(\Gm^{A_1\cdots A_6}C^{-1})^{\al\beta}
{}^{(\mp)}Z^{(6)}_{A_1\cdots A_6} \ ,
\label{wp}
\ee
where $T^{(\mp)}_{AB} = Z^{(2)}_{AB}\mp \ast Z^{(10)}_{AB}$ and 
$\ast Z^{(10)}_{AB}$ is the (10+2)D dual of $Z^{(10)}$.
The tensor $Z^{(6)}$ becomes (anti-)self-dual under the projection.
This is the $N=(1,0)$ superalgebra in the (10+2)D flat spacetime,
with 32 real supercharges \cite{ib},\cite{hp}.
\par

Let us consider the dimensional reduction to (10+1)D by the 
compactification of the $\t0$ time direction.
We choose the projection $P_-$ with the explicit realization 
(\ref{gm1}), (\ref{gm2}) of gamma matrices, which leads to the expression 
$Q_-^{\al} = (0, Q^{\tal})$ $(\tal = 1,\cdots,32)$.
With the redefinition of variables, $P_a = -T^{(+)}_{a \t0}$, 
$Y_{ab}=-T^{(+)}_{ab}$,
$Z_{a_1 \cdots a_5} = 2{}^{(+)}Z^{(6)}_{\t0 a_1\cdots a_5}$, (\ref{wp}) 
becomes
\be
\{Q^{\tal}, Q^{\tbe}\} = (\gm^a C_{(11)}^{-1})^{\tal\tbe} P_a 
+ \hf (\gm^{a b}C_{(11)}^{-1})^{\tal \tbe}Y_{ab} 
+ {1 \over 5!}(\gm^{a_1 \cdots a_5}C_{(11)}^{-1})^{\tal \tbe} 
Z_{a_1 \cdots a_5} \ ,
\label{wp2}
\ee
where $C_{(11)} = \gm^0$ is the charge conjugation matrix in (10+1)D 
and $Q^{\tal}$ is the (10+1)D Majorana spinor with respect to $C_{(11)}$.
This is the (10+1)D M-theory superalgebra \cite{pt2}.  
\par

We further take the $S^1$ compactification of the spatial 10-direction 
of the (10+1)D spacetime.
We decompose $Q^{\tal}$ into two Majorana-Weyl supercharges 
$Q_{A\pm}^{\tal}$ of opposite chirality with respect to the 
matrix $\gm^{10}$.
Acting the projector $P^{(10)}_{\pm} = \hf (I_{32}\pm \gm^{10})$
on (\ref{wp2}), we obtain the (9+1)D Type IIA superalgebra, 
\bea 
\{Q_{A\pm}^{\tal}, Q_{A\pm}^{\tbe}\} = &&
(P^{(10)}_{\pm}\gm^{\mu}C_{(10)}^{-1})^{\tal\tbe} (P_{\mu} \mp Z_{\mu})
+ {1 \over 5!} (\gm^{\mu_1 \cdots \mu_5}C_{(10)}^{-1})^{\tal \tbe}
{}^{(\mp)}Z_{\mu_1 \cdots \mu_5} \ , \nonumber \\
\{Q_{A\pm}^{\tal}, Q_{A\mp}^{\tbe}\} = &&
\pm (P^{(10)}_{\pm}C_{(10)}^{-1})^{\tal \tbe}Z 
+\hf (P^{(10)}_{\pm}\gm^{\mu \nu}C_{(10)}^{-1})^{\tal \tbe}Y_{\mu\nu}
\\
&& \pm {1 \over 4!} 
(P^{(10)}_{\pm} \gm^{\mu_1 \cdots \mu_4}C_{(10)}^{-1})^{\tal \tbe}
Z_{\mu_1 \cdots \mu_4} \ , \nonumber 
\label{IIA}
\eea 
where $Z = P_{10}$, $Z_{\mu} = Y_{\mu10}$, 
$Z_{\mu_1 \cdots \mu_4}= Z_{10\mu_1 \cdots \mu_4}$ and 
$C_{(10)} = C_{(11)} = \gm^0$.

\subsection{IIB projection}%
Let us take another projection of (\ref{qq}) by the projector
$P_{B\pm} = \hf (I_{64}\pm {\hat \Gm})$ with 
${\hat \Gm} \equiv \Gm^0\Gm^1\cdots \Gm^9$ \cite{hp}.
The supercharges 
$Q_{B\pm}^{\al} = (P_{B\pm})^{\al}_{\ \beta}Q^{\beta}$ are 
not $SO(10,2)$ covariant but covariant under $SO(9,1)$.
In this case, all terms in the RHS of (\ref{qq}) remain after the 
projection.
We choose $P_{B+}$ with the representation (\ref{gm1}), (\ref{gm2}), which
becomes
\be
P_{B+} = \left(
\begin{array}{cc} P^{(10)}_+ \ & 0 \ \\ 0 \ & P^{(10)}_+ \ 
\end{array} \right) \ .
\ee
The projected $Q_{B+}^{\al}$ are written as
$Q_{B+}^{\al} = (Q^{1 \tal}_B, Q^{2 \tbe}_B)$, where two $Q^{i \tal}_B$ have 
the same chirality with respect to $\gm^{10}$. 
Then, compactifying timelike and spacelike $(\t0,10)$-directions, we obtain the 
(9+1)D Type IIB superalgebra,
\bea 
\{Q^{i \tal}_B, Q^{j \tbe}_B\} =
&&\delta^{ij}(P^{(10)}_+ \gm^{\mu}C_{(10)}^{-1})^{\tal \tbe}
(P_{\mu} - T_{\mu})
+ (P^{(10)}_+\gm^{\mu}C_{(10)}^{-1})^{\tal \tbe}{\tilde Z}^{ij}_{\mu}
\nonumber \\
&&+ {1 \over 3!} \epsilon^{ij}
(P^{(10)}_+ \gm^{\mu_1\mu_2\mu_3}C_{(10)}^{-1})^{\tal \tbe}
T_{\mu_1\mu_2\mu_3} \nonumber
+ {1 \over 5!} \delta^{ij}
(\gm^{\mu_1 \cdots \mu_5}C_{(10)}^{-1})^{\tal \tbe}
{}^{(-)}Z_{\mu_1 \cdots \mu_5} \\
&&+ {1 \over 5!}(\gm^{\mu_1 \cdots \mu_5}C_{(10)}^{-1})^{\tal \tbe}
{}^{(-)}{\tilde Z}^{ij}_{\mu_1 \cdots \mu_5} \ , 
\label{IIB}
\eea 
where $P_{\mu}=Z^{(2)}_{\t0\mu}$, $T_{\mu}= \ast Z^{(10)}_{10\mu}$,
$Z_{\mu_1 \cdots \mu_5} = Z^{(6)}_{\t0\mu_1\cdots\mu_5}$ and 
\bea 
&&{\tilde Z}^{ij}_{\mu} = (\sg_3)^{ij}(Z^{(2)}_{10\mu}-\ast Z^{(10)}_{\t0\mu}) 
+(\sg_1)^{ij}(\ast Z^{(11)}_{\mu} - Z^{(3)}_{\t0 10 \mu}) \ , \\
&&T_{\mu_1\mu_2\mu_3} = \ast Z^{(7)}_{\t0 10 \mu_1\mu_2\mu_3}
-Z^{(3)}_{\mu_1\mu_2\mu_3}  \ , \quad
{\tilde Z}^{ij}_{\mu_1 \cdots \mu_5} = 
(\sg_3)^{ij}Z^{(6)}_{10\mu_1\cdots\mu_5}
+(\sg_1)^{ij}\ast Z^{(7)}_{\mu_1\cdots\mu_5} \ .
\nonumber
\eea 
\par

\vspace{0.3cm}
Note that the positivity condition $Z^{(2)}_{\t0 0} \geq 0$ in (\ref{pos}) 
is translated into $P_0 \geq 0$ under the above two projections.
Having clarified the relation of the (10+2)D $N=(1,1)$ algebra
to M-, type IIA and IIB algebras, in the next section we investigate 
the connection of BPS states inherent in the (10+2)D algebra with those in 
M-theory and string theories, starting from the classification of 
the (10+2)D BPS states.
\par

\section{(10+2)D BPS states}%
\subsection{BPS states in (10+1)D M-algebra}%
Before classifying BPS states in (10+2)D, it is better to recall 
how to find BPS states in an ordinary Poincar\'e superalgebra.
As an example, let us take the M-algebra in (10+1)D, with only 
the first term in the RHS of (\ref{wp2}),
\be
\{Q^{\tal}, Q^{\tbe}\} = (\gm^a C_{(11)}^{-1})^{\tal \tbe} P_a \ .
\label{sp}
\ee
A BPS state is annihilated by some combination of supercharges
and hence has a configuration with ${\rm det}\{Q^{\al}, Q^{\beta}\}=0$, 
which is equivalent to the null condition $P^2 = 0$.
If $P_a$ is null, by the action of the $SO(10,1)$ rotation, we can choose
a frame in which $P_a = (\lm, \pm \lm, 0, \cdots ,0)$, with a positive 
parameter $\lm$. 
Then (\ref{sp}) becomes 
\be
\{Q^{\tal}, Q^{\tbe}\} = \lm (I \pm \gm_1\gm_0)^{\tal \tbe} \ .
\ee
Since the matrix $\gm_1\gm_0$ squares to the identity and is also traceless,
the half of eigenvalues of $\gm_1\gm_0$ are $+1$ and half $-1$.
Thus we see that there exists a unique BPS state in (\ref{sp}) breaking 
half of the supersymmetries, which corresponds to a massless 
particle in (10+1)D \cite{pt2}. 
We need the second and third central terms in the RHS of (\ref{wp2}) 
to obtain other BPS states, e.g. M2 and M5 branes.

\subsection{BPS states in (10+2)D superalgebra}%
\subsubsection{Second-rank central term $Z_{AB}$}
We now consider what types of BPS states exist in the (10+2)D 
$N=(1,1)$ superalgebra.
In this paper, we concentrate on the anti-symmetric central term 
$Z^{(2)}_{AB}$ in the RHS of (\ref{qq}) and make other $Z^{(k)}$ zero;
\be
\{Q^{\al}, Q^{\beta}\} = 
{1 \over 2} (\Gm^{AB}C^{-1})^{\al\beta} Z^{(2)}_{AB} \ .
\label{qqz}
\ee
In order to find possible BPS configurations inherent in 
(\ref{qqz}), we intend to simplify the form of $Z^{(2)}_{AB}$
by the action of the $SO(10,2)$ group, just as the case of the translation 
generator $P_a$ in the above M-algebra.
The matrix $Z = (Z^{(2) A}_{\ \ \ \ \, B})$ may be regarded as a map of the
12D vector space $V_{(12)}$ onto $V_{(12)}$; for $v=(v^A) \in V_{(12)}$, 
$Zv = (Z^{(2) A}_{\ \ \ \ \, B}v^B) \in V_{(12)}$.
We see that $Z$ belongs to the $so(10,2)$ algebra represented on
$V_{(12)}$,
\be
so(10,2) = \{Z \in gl(12, R); \ \eta(Zu,v) + \eta(u,Zv) = 0 \, , 
\ {\rm for}\ {\rm all}\ u,v \in V_{(12)} \} \ ,
\label{so}
\ee
where $\eta$ is the (10+2)D metric $\eta(u,v) = \eta_{AB}u^Av^B$.
An arbitrary element $Z \in SO(10,2)$ is transformed under the adjoint action 
of $SO(10,2)$; $Z \rightarrow \Lambda^{-1} Z \Lambda$, $\Lambda \in SO(10,2)$.
The problem to find BPS states then reduces, as a first step, to that of  
classifying all conjugacy classes in the $so(10,2)$ algebra under the adjoint 
action.
In 12D Euclidean space-time, under the $SO(12)$ rotation, any 
$Z$ in the $so(10,2)$ algebra is reduced to canonical form which has six 
$2 \times 2$ anti-symmetric blocks $A_k = i h_k \sg_2$ $(k=1,\cdots,6)$ 
on the diagonal part.
However, in pseudo-Euclidean space-time, some configurations of $Z$ can not be 
brought to canonical form and we have to be careful for the classification. 
\par

It is known that an arbitrary element $Z$ in a semisimple Lie algebra 
{\eufm g} has a unique decomposition,
\be
Z = S + N \ , \qquad [S, N] = 0 \ .
\ee
where $S$ and $N$ are semisimple and nilpotent elements in {\eufm g}, 
respectively.
\par

An element $S \in$ {\eufm g} is called semisimple if its adjoint 
representation $ad(S)$ is a diagonalizable matrix. 
In complex or real compact Lie algebras, any semisimple element 
is reduced to a unique form like the above $SO(12)$ case, since there
exists a unique Cartan subalgebra in {\eufm g}.
In the real non-compact case, however, there are 
several non-equivalent Cartan subalgebras under the adjoint group
$G$.
An example is the $sl(2,R)$ algebra, which has two Cartan algebras, 
$\{\lm \sg_3\}$ and $\{i \theta \sg_2\}$ $(\lm, \theta \in R)$. 
These are obviously inequivalent, since the former generates a 
non-compact group, while the latter yields the 2D rotation.
The classification of all Cartan subalgebras in real semisimple
Lie algebras is given in a paper of Sugiura \cite{ms}.
In the real non-compact case, there also appears a nilpotent element 
$N \in$ {\eufm g} such that $N^m \not=0$ and $N^{m+1}=0$ for 
some non-negative integer $m$.
Thus, in our superalgebra defined on the pseudo-Euclidean spacetime
with two time directions, we expect to obtain several conjugacy classes 
in $so(10,2)$ containing nilpotent elements.
\par

The complete classification of real semisimple Lie algebras under the 
action of its adjoint group is given in a paper of Burgoyne and Cushman
\cite{bc}. 
In the following, we classify the $so(10,2)$ algebra according to the 
method in \cite{bc} and construct a representative for each conjugacy 
class in $so(10,2)$.

\subsubsection{Types $\dlt$}
We give a brief review on results in \cite{bc} which are needed 
in the rest of this paper.
We start with the general linear group $GL(12,C)=GL(V_{(12)})$ 
represented on the 12D complex vector space $V_{(12)}$.
The orthogonal group $O(10,2)$ is defined as a real form of 
$GL(12,C)$ specified by an automorphism $\sg$ of $GL(12,C)$ with 
$\sg^2 =+1$,
\be
O(10,2) = \{g \in GL(12,C) \, | \, \eta(gu,gv) = \eta(u,v) \ {\rm for}\ 
{\rm all}\ u,v \in V_{(12)} \ , \ \sg^{-1} g \sg = g \} \ .
\ee
Note that $\sg$ acts on $V_{(12)}$ as an anti-linear map onto 
$V_{(12)}$ with $\eta(\sg u, \sg v) = \eta(u,v)^{\ast}$.
The corresponding Lie algebra of $O(10,2)$ is  
$o(10,2) = o(V_{(12)},\eta, \sg) \simeq so(10,2)$, which 
is given in (\ref{so}).
\par

We introduce an equivalence relation among pairs of the form
$(Z,V_{(n)})$, where $Z \in o(V,\tau_{(p)},\sg)$.
The algebra $o(V,\tau_{(p)},\sg)$ is associated with the group 
$O(n-p,p)$ and represented on $V_{(n)}$ with the metric $\tau_{(p)}$.
Let $Z' \in o(V'_{(n)},\tau'_{(p)}, \sg')$ then we write 
$(Z, V_{(n)}) \sim (Z', V'_{(n)})$ if there exists an isomorphism $\phi$ 
of $V_{(n)}$ onto $V'_{(n)}$ such that $\phi Z = Z' \phi$,
$\phi \sg = \sg' \phi$ and $\tau_{(p)}(u,v) = \tau'_{(p)}(\phi u, \phi v)$
for $u,v \in V_{(n)}$.
In \cite{bc}, an equivalence class for $\sim$ is called a {\it type}
$\dlt$ and its dimension is given by $dim.\dlt = dim.V_{(n)}=n$ if 
$(Z, V_{(n)}) \in \dlt$.
The index of $\dlt$ denoted as $ind.\dlt$ is the number of time 
(negative sign) components in the metric $\tau_{(p)}$ in 
$o(V_{(n)},\tau_{(p)},\sg)$, i.e. $ind.\dlt = p$.
The type $\dlt$ is nothing but a conjugacy class of the 
$o(n-p,p)$ algebra under the adjoint action of $O(n-p,p)$;
\par
\vspace{0.4cm}
\noindent
Proposition 1. (\cite{bc}, Sec.\,2.1) \\
{\it Let $Z, X \in o(V_{(n)},\tau_{(p)},\sg)$. There exists a 
$g \in O(V_{(n)},\tau_{(p)},\sg)$ such that $g^{-1}Zg = X$\\
if and only if $(Z, V_{(n)})$ and $(X, V_{(n)})$ belong to the same type.}
\par
\vspace{0.4cm}

Next, we introduce the notion of indecomposable type. 
Let $Z \in o(V_{(n)}, \tau_{(p)}, \sg)$ and let $\dlt$ denote the type
containing $(Z,V_{(n)})$.
We suppose that $V_{(n)} = W_1 + W_2$ is a sum of proper, disjoint,
$Z$-invariant, $\sg$-invariant, and orthogonal subspaces.
Let $\tau_{(p)}^i$ $(i=1,2)$ be the restriction of the metric $\tau_{(p)}$ 
to each $W_i$, then the restriction of $Z$ to each $W_i$ denoted as
$Z_i$ belongs to $o(W_i,\tau_{(p)}^i,\sg)$.
Let $\dlt_i$ denote the type containing $(Z_i, W_i)$.
Then we write $\dlt = \dlt_1 + \dlt_2$.
A type $\dlt$ is called ${\it indecomposable}$ if it can not be
written as the sum of two or more types.
For any type $\dlt$, we have
\par
\vspace{0.4cm}
\noindent
{\bf Theorem.} (\cite{bc}, Sec.\,2.2)\\
{\it The decomposition $\dlt = \dlt_1 + \dlt_2 + \cdots + \dlt_s$ into 
indecomposable types is unique.}
\par
\vspace{0.4cm}
\noindent
In the decomposition, a representative $Z$ in $\dlt$ can be given as a matrix 
to have representatives $Z_k$ in $\dlt_k$ $(k=1,\cdots,s)$ on 
its diagonal part, with respect to the metric $\tau_{(p)}$ which has metrices
$\tau_{(p)}^k$ in $\dlt_k$ on its diagonal.
If a type $\dlt$ is decomposed as above, for the dimension and the index 
of $\dlt$, we have 
\bea
&& dim.\dlt = dim.\dlt_1 + dim.\dlt_2 + \cdots + dim.\dlt_s \ , 
\nonumber \\
&& ind.\dlt \  = ind.\dlt_1 \  + ind.\dlt_2 \  + \cdots + ind.\dlt_s \ .
\label{d-i}
\eea
In Table 2 below, we write down all indecomposable types and 
their dimension and index for the $o(q,p)$ algebra.
In our $O(10,2)$ case, all types $\dlt$, i.e., all conjugacy 
classes in the $o(10,2)$ algebra can be obtained by taking all 
possible combinations of the indecomposable types with the conditions 
$dim.\dlt=12$ and $ind.\dlt=2$, which will be done in the following
sections.
\par
\newpage
\begin{center} Table 2. Indecomposable types in the $o(q,p)$ algebra\\
\begin{tabular}{|cllcc|} \hline
Type & & & $dim.\dlt$ & $ind.\dlt$ \\ \hline
$\dlt_m(\zt,-\zt,{\bar \zt}, -{\bar \zt})$
&$\zt \not= \pm {\bar \zt}$ &  & $4(m+1)$ & $2(m+1)$ \\
$\dlt_m(\zt, -\zt)$ &$\zt={\bar \zt}\not=0$ & & $2(m+1)$ & $m+1$ \\
$\dlt_m^{\ep}(\zt,-\zt)$ &$\zt = - {\bar \zt} \not=0$ & $m$: even &
$2(m+1)$ & $m+1 - (-1)^{m \over 2}\ep$ \\
& & $m$: odd & $2(m+1)$ & $m+1$ \\
$\dlt_m^{\ep}(0)$ & & $m$: even & $m+1$ & 
$\hf (m+1 - (-1)^{m \over 2}\ep)$ \\
$\dlt_m(0,0)$ & & $m$: odd & $2(m+1)$ & $m+1$ \\ \hline
\end{tabular}
\end{center}
\par

We briefly explain the notation of indecomposable types in Table 2
and how to construct a representative for each type.
Henceforth we omit the subscripts of $V_{(n)}$ and $\tau_{(p)}$
for simplicity.
Let $(Z,V) \in \dlt_m(\zt_1,\cdots,\zt_t)$, where the subscript 
$m = 0,1,\cdots$ means that, in the decomposition $Z = S + N$ in 
Sec.\,3.2.1, the nilpotent $N$ satisfies $N^m \not=0$ and $N^{m+1} = 0$, 
while parameters $\zt_1,\cdots,\zt_t \in C$ are eigenvalues of the 
semisimple element $S$.
Let $F$ be a subspace of the vector space $V$ spanned by eigenvectors 
$\{e_1,\cdots,e_t\}$ of $S$.
We can show that $V$ is decomposed as $V = F + NF + \cdots + N^m F$.
Now the semisimple $S$ can be supposed to belong to the 
algebra $o(F,\tau_m,\sg)$ represented on $F$ with the metric 
$\tau_m(u,v) \equiv \tau(u, N^mv)$ for $u,v \in F$.
Note that, from the nilpotency of $N$, the metric $\tau_m(u,v) = 0$ 
for $u, v \in NV$.
Non-zero values of $\tau_m$ are given for each indecomposable type in 
the Appendix 2 of \cite{bc}.
The description of the indecomposable types enables us to construct an 
explicit matrix realization of $Z = S + N \in o(V,\tau,\sg)$ and the 
metric $\tau$ on $V$.
In particular, descriptions of $Z$ and $\tau$ as real matrices on $V$
can be given easily by taking $\sg$-invariant combinations of elements
in $V$.
In the Appendix in this paper, we give explicit forms of representatives 
for all indecomposable types relevant to our $O(10,2)$ case.
Having the representatives and arranging them on the diagonal part 
of $12 \times 12$ matrix, we obtain a realization of a representative
for each type in $o(10,2)$, which is with respect to a (10+2)D metric 
with metrices of indecomposable types arranged on its diagonal part.
Finally, by an isomorphism $\phi$ of $V$ onto $V$ noted above, we 
have our seeking forms of representatives with respect to our (10+2)D 
metric $\eta = diag.(-,-,+\cdots,+)$.
\par

As a final step, we move to the classification of the $so(10,2)$ 
algebra under the action of the $SO(10,2)$ group, rather than $O(10,2)$.
More exactly, we classify $so(10,2)$ under the subgroup of $SO(10,2)$
generated by the $so(10,2)$ generators $M_{AB}$, which we denote as 
$SO(10,2)_0$.
The subgroup is the connected part of $SO(10,2)$ containing the identity
element and its action to an arbitrary element $Z$ in $so(10,2)$ does not
change the sign of the component $Z_{\t0 0}$ in (\ref{pos}).
The quotient group $O(10,2)/SO(10,2)_0$ is equal to $Z_2 \times Z_2$.
One of $Z_2$'s corresponds to the sign of determinant, while 
another $Z_2$ indicates that $SO(10,2)$, or more generally $SO(q,p)$, 
has two connected parts.
The two connected parts are distinguished by the sign of the 
principal minor of the time components of $\Lambda \in SO(q,p)$ 
\cite{ov}.
In $SO(10,2)$, we have the condition
$(\Lambda^{\t0}_{\ \t0} \Lambda^0_{\ 0} - 
\Lambda^0_{\ \t0}\Lambda^{\t0}_{\ 0})^2 \geq 1$, where
the positive sign of the determinant corresponds to $SO(10,2)_0$.
We choose an element $D$ ($C$) in $O(10,2)$ which has
the negative (positive) determinant and the positive (negative)
principal minor for the time components.
If the action of $C$ and $D$ on a representative $Z$ in a type $\dlt$ 
in $O(10,2)$,
\be
{}^{(+,+)}Z = Z \ , \ \ 
{}^{(+,-)}Z = C^{-1}Z \, C \ , \ \ 
{}^{(-,+)}Z = D^{-1}Z \, D \ , \ \ 
{}^{(-,-)}Z = (CD)^{-1}Z \, (CD) \ ,
\ee
is non-trivial, then the type $\dlt$ splits into four distinct types 
${}^{(\ep_D,\ep_C)}\dlt \ni {}^{(\ep_D,\ep_C)}Z$, $(\ep_D, \ep_C = \pm1)$ 
under $SO(10,2)_0$.
Some $\dlt$ may split into only two types ${}^{(\ep_D,+)}\dlt$ 
(${}^{(+,\ep_C)}\dlt$) if $Z$ in $\dlt$ is invariant
under the $C$ ($D$) action, while some $\dlt$ may be invariant under 
$SO(10,2)_0$.
\par

With the description noted above, we give all types $\dlt$, i.e., 
conjugacy classes of the $so(10,2)$ algebra under the $SO(10,2)_0$ 
action and find possible BPS configurations coming from the types.
\par

\subsection{$so(2,2)$ case}%
\subsubsection{$so(2,2)$ conjugacy classes}
Before studying the full $so(10,2)$ case, we demonstrate the 
classification of its $so(2,2)$ part.
We will see that almost possible BPS states arise from this part.
For the $SO(2,2)$ group, it is known that its associated Lie algebra 
has the decomposition $so(2,2) \simeq sl(2,R)_L \times sl(2,R)_R$.
We take the basis $\{H_i, E_i, F_i\}$ for each $sl(2,R)_i$ satisfying
\be
[H_i, E_i] = E_i \ , \quad  [H_i, F_i] = - F_i \ , \quad
[E_i, F_i] = 2 H_i \ ,
\ee
with an explicit realization of them,
\be
H_L = \hf \left( \begin{array}{cc}0 \ & I_2 \ \\ I_2 \ & 0 \ 
\end{array} \right) \ , \qquad
H_R = \hf \left(\begin{array}{cc}0 \ & \sg_3 \ \\ \sg_3 \ & 0 \ 
\end{array} \right) \ , 
\ee
and 
\be
E_L = \hf \left(\begin{array}{cc} \ve_2 \ & -\ve_2 \ \\ \ve_2 \ & 
-\ve_2 \ \end{array} \right)  \ , \qquad
E_R = \hf \left(\begin{array}{cc} \ve_2 \ & -\sg_1 \ \\ -\sg_1 \ & 
\ve_2 \  \end{array} \right) \ ,
\ee
and $F_L = {\tilde E}_L$ and $F_R = {\tilde E}_R$, where the $2\times2$ real 
matrix $\ve_2 = -i \sg_2$.
For later use, we define
\be
K_L = E_L - F_L = 
\left( \begin{array}{cc}\ve_2 \ & 0 \ \\ 0 \ & -\ve_2 \ 
\end{array} \right) \ , \qquad
K_R = E_R - F_R = 
\left(\begin{array}{cc}\ve_2 \ & 0 \ \\ 0 \ & \ve_2 \ 
\end{array} \right) \ .
\ee
For the $C$- and $D$-splittings noted in Sec.\,3.2.2, we use
\be
C = 
\left( \begin{array}{cc}\sg_3 \ & 0 \ \\ 0 \ & \sg_3 \ 
\end{array} \right) \ , \qquad
D = 
\left( \begin{array}{cc} I_2 \ & 0 \ \\ 0 \ & \sg_3 \ 
\end{array} \right) \ . \qquad
\ee
In the following, we show representatives for all conjugacy classes 
of the $so(2,2)$ algebra with respect to the (2+2)D metric 
$\eta_{AB} = (-,-,+,+)$, $(A,B = \t0,0,9,10)$ in terms of 
$\{H_i, E_i, F_i\}$ and $K_i$. 
\\
(I) Semisimple  cases
\bea
&&(i) \ {\cal S}_1 = (h_1+h_2) H_1 + (h_1-h_2) H_2 \ 
\in {}^{(\pm,+)}(\dlt_0(h_1,-h_1) + \dlt_0(h_2,-h_2)) \ , \nonumber \\
&&\hspace*{9.0cm}
(h_1, h_2 \in R, \ h_1 \geq 0, \ |h_2| \leq 0)
\\ 
&&(ii) \ {\cal S}_2 =\hf (h_1+h_2) K_L + \hf (h_1-h_2) K_R \ 
\in {}^{(\pm,\pm)}(\dlt^-_0(-i h_1,i h_1) + \dlt^+_0(i h_2, -i h_2))
\ , \nonumber \\
&&\hspace*{9.0cm}   (h_1, h_2 \in R) \\
&&(iii) \ {\cal S}_3 = 2 h_1 H_L + h_2 K_R \ 
\in {}^{(+,\pm)}\dlt_0(\zt,-\zt,{\bar \zt}, -{\bar \zt}) \ , \nonumber \\
&&\hspace*{6.5cm}
(\zt = h_1 + i h_2, \ h_1 \geq 0, \ h_2 \in R) \\
&&(iv) \ {\cal S}_4 = h_2 K_L + 2 h_1 H_R \ 
\in {}^{(-,\pm)}\dlt_0(\zt,-\zt,{\bar \zt},-{\bar \zt})  \ , \nonumber \\
&&\hspace*{6.5cm}
(\zt = h_1 + i h_2, \ h_1 \geq 0, \ h_2 \in R) \ .
\eea
In (I-{\it i}), $\ep_D=+1$ ($-1$) corresponds to the positive (negative) 
$h_2$.
In (I-{\it ii}), four distinct types labeled by $(\pm,\pm)$ correspond to four 
quadrants in the $(h_1,h_2)$ plane.
The sign $\ep_C=+1$ ($-1$) is for the positive (negative) $h_2$ in 
(I-{\it iii}) and (I-{\it iv}).
\par
\vspace{0.4cm}
\noindent
(II) Nilpotent cases
\bea
&& (i) \ {\cal N}_{1\pm} = \pm (E_L + E_R) \ 
\in {}^{(+,\pm)}(\dlt^+_2(0) + \dlt^+_0(0)) \ , \\
&&(ii) \ {\cal N}_{2\pm} = \pm (E_L - E_R) 
\ \in {}^{(\mp,+)}(\dlt^-_0(0) + \dlt^-_2(0)) \ , \\
&&(iii) \ {\cal N}_{3\pm}  = \pm E_L \ \in {}^{(+,\pm)} \dlt_1(0,0) 
\ , \\
&&(iv) \ {\cal N}_{4\pm} = \pm E_R \ \in {}^{(-,\pm)}\dlt_1(0,0) \ .
\eea
\par
\vspace{0.4cm}
\noindent
(III) Semisimple $+$ nilpotent cases
\bea
&&(i) \ {\cal M}_{1\pm} = \ \ \ \, h H_L \pm E_R \ \in 
{}^{(-,\pm)}\dlt_1(-h/2,h/2) \ ,  \quad \, (h > 0) \\
&&(ii) \ {\cal M}_{2\pm} = \pm E_L + h H_R \ \in 
{}^{(+,\pm)}\dlt_1(-h/2,h/2) \ , \, \quad (h > 0) \\
&&(iii) \ {\cal M}_{3\pm} = \pm E_L + h K_R \ \in 
{}^{(-,\mp)}\dlt^{+}_1(i h, -i h) \ , \qquad (h > 0) 
\nonumber \\
&&\hspace*{4.7cm} \, \in {}^{(+,\mp)}\dlt^-_1(i h, -i h) \ , 
\quad \ \ \ (h < 0) \\
&&(iv) \ {\cal M}_{4\pm} = h K_L \pm E_R \ \in 
{}^{(+,\mp)}\dlt^{+}_1(i h, -i h) \ , \quad \ \ \ \ \ \, (h > 0) 
\nonumber \\
&&\hspace*{4.4cm} \in {}^{(+,\mp)}\dlt^-_1(i h, -i h) \ , 
\quad \ \ \ \ \ \, (h < 0)  \ .
\eea
\par

\subsubsection{BPS states}
Having obtained all inequivalent representatives in $so(2,2)$, 
we substitute them into the RHS of (\ref{qqz}) and calculate the
characteristic polynomial,
\be
D(x) = \det.(\{Q,Q\} - x I_{64}) \ , 
\ee
for the $64 \times 64$ matrix $\{Q^{\al},Q^{\beta}\}$.
We also denote the characteristic polynomial for the Weyl-projected
$32 \times 32$ matrix $\{Q^{\tal}, Q^{\tbe}\}$ as $D_W(x)$.
Note that eigenvalues of $\{Q,Q\}$ do not depend on the choice of
a particular representative of types in $so(2,2)$. 
As mentioned in Sec.\,2, by the positivity condition, the eigenvalues must 
be non-negative. 
In the following, we show only the cases which satisfy the positivity 
condition.
\par

\vspace{0.5cm}
\noindent 
[I] ${\cal S}_2 = \hf (h_1+h_2) K_L + \hf (h_1-h_2) K_R$ \\
With the form of ${\cal S}_2$, (\ref{qqz}) becomes
\be
\{Q^{\al}, Q^{\beta}\} = h_1I_{64} - h_2\Gm^{\t0 0 9 10} \ ,
\ee
with its characteristic polynomial,
\be
D(x) = (D_W(x))^2 \ , \qquad D_W(x) = 
(x-(h_1+h_2))^{16} (x - (h_1 - h_2))^{16} \ ,
\ee
where $h_1 = Z_{\t0 0} \geq 0$.
The bound $h_1 \geq |h_2|$ is imposed by the positivity condition.
A BPS state arises if and only if the bound is saturated, i.e., 
$h_1 = |h_2|$, in which the fraction of survived supersymmetries 
is $1/2$.
In this case, the eigenspinor $\ep$ of $\{Q^{\al}, Q^{\beta}\}$ with
zero eigenvalue satisfies
\be
\Gm^{\t0 0 9 10} \ep = \pm \ep \ ,
\label{22b}
\ee
where the sign $+(-)$ is for positive (negative) $h_2$.
{}From (\ref{22b}), it is natural to interpret 
the ${\cal S}_2$ case as the configuration of an extended object in 
the $(\t0,0,9,10)$ directions, which we call super (2+2)-brane.
We note that ${\cal S}_3$ and ${\cal S}_4$ cases with the positivity
condition reduce to the BPS saturated configuration in the ${\cal S}_2$
case.
\par

Let us check how the (2+2)-brane is reduced to other branes in M-theory 
and type IIA theory by the dimensional reduction described in Sec.\,2.3.
The condition (\ref{22b}) with, say, the minus sign in the RHS becomes 
under the Weyl projection,
\be
\gm^{0 9 10} \ep_- = \ep_- \ ,
\label{mb}
\ee
where $\ep_-$ is a chiral part of $\ep$ with $\Gm \ep_- = - \ep_-$.
(\ref{mb}) shows that the dimensionally reduced super (2+2)-brane
is nothing but a supermembrane extended in the $(0,9,10)$ directions.
The further dimensional reduction of the spatial $10$-direction yields the 
type IIA fundamental string configuration in the $(0,9)$ directions, 
characterized by the condition
\be
\gm^{09} \ep_L = \ep_L \ , \qquad 
\gm^{09} \ep_R = - \ep_R \ ,
\label{asb}
\ee
where $\ep_{L(R)}$ is a chiral part of $\ep_-$ with chirality 
$+1$ $(-1)$ with respect to the (9+1)D chirality matrix $\gm^{10}$.
\par

On the other hand, under dimensional reduction to (9+1)D with
the IIB projection in Sec.\,2.4, (\ref{22b}) becomes 
\be
\gm^{09} \ep_L = \ep_L \ , \qquad
\gm^{09} \ep_R = - \ep_R \ ,
\label{bsb}
\ee
where the eigenspinor $\ep$ is decomposed as $\ep =(\ep_R, \ep_L)$ with
$\gm^{10}\ep_{L(R)} = \ep_{L(R)}$.
It is known that (\ref{bsb}) can be derived from (\ref{asb}), upon
compactification to (8+1)D, by the T-duality transformation
and corresponds to the (9+1)D type IIB fundamental string. 
\par

\vspace{0.5cm}
\noindent
[II] ${\cal N}_{1+} = E_L + E_R$ \\
In this nilpotent case, $Z_{\t0 0}= Z_{09} =1$ and the others zero.
Then (\ref{qqz}) takes the form,
\be
\{Q^{\al}, Q^{\beta}\} = I_{64} + \Gm^{9\t0} \ ,
\ee
with 
\be
D(x) = (D_W(x))^2 \ , \qquad
D_W(x) = x^{16}(x-2)^{16} \ .
\label{n1}
\ee
This ${\cal N}_{1+}$ case also yields a BPS state with $1/2$ 
surviving supersymmetries. 
The BPS state, however, seems rather different from the one in the 
${\cal S}_2$ case, since, from (\ref{n1}), the former is definitely 
BPS saturated without any non-BPS excited configuration.
It is caused by the fact that the nilpotent $Z_{AB}$ is parameter 
independent.
\par

Let us see how the BPS state is observed after the dimensional reduction 
to M-theory. 
(\ref{n1}) indicates that the reduced state in M-theory is still 
BPS with $1/2$-susy, which has a non-zero $Y_{09}$ charge in 
the algebra (\ref{wp2}).
In \cite{ch},\cite{pt2}, it was argued that the dual of $Y_{09}$ 
corresponds to a charge of the M9 brane extended in $(0,1,\cdots,8,10)$
directions.
Subsequently, in \cite{bs}, the target space solution for the M9 brane
was found to be a domain wall solution of the massive (10+1)D supergravity.
\par

Under another dimensional reduction to (9+1)D, the (10+2)D BPS state does not
reduce to any BPS state in the type IIB theory.
In this reduction, $Z_{\t0 0}$ becomes the energy $P_0$ in (9+1)D, while 
$Z_{09}$ does not appear in the algebra (\ref{IIB}).
Hence the reduced state is interpreted as a non-BPS massive particle
in the type IIB theory.
\par

\vspace{0.5cm}
\noindent
[III] ${\cal M}_{3+}= E_L + h K_R$ \\
The characteristic polynomial in this semisimple$+$nilpotent case becomes
\be
D(x) = (D_W(x))^2 \ , \qquad
D_W(x) = x^8 (x-2 h)^{16} (x - 2)^8 \ .
\ee
The eigenvalues of the matrix $\{Q,Q\}$ are $x = 0, 2, 2 h$ and thus the 
parameter $h$ must be positive.
As the number of zero eigenvalues is 8, this ${\cal M}_{3+}$ case
corresponds to a BPS state with $1/4$-susy.
In the case of $h=0$, which is the nilpotent ${\cal N}_{3+}$ case, 
there arise 16 additional zero eigenvalues and the fraction of surviving 
supersymmetries becomes $3/4$.
\par

These fractions $1/4$ and $3/4$ remain unchanged under the Weyl projection.
The ${\cal M}_{3+}$ case reduces to the $1/4$-susy configuration with 
$P_0 = h+1/2$, $P_{10}= -1/2$, $Y_{09}= -1/2$, $Y_{910} = h-1/2$ and
the others zero in the M-algebra (\ref{wp2}).
Since, as noted in [I],\,[II], $Y_{910}$ and $Y_{09}$ are identified 
with M2 and M9 brane charges, it would be possible to interpret the 
configuration as a composite state of these branes.
It is needed to investigate whether the interpretation is also applicable 
to the ${\cal N}_{3+}$ case, or not.
\par

Under the type IIB projection, the ${\cal M}_{3+}$ case reduces to the 
IIB state with $P_0 = h+1/2$, 
${\tilde Z}^{11}_9 = - {\tilde Z}^{22}_9 = h-1/2$ and the others zero 
in the algebra (\ref{IIB}).
It is obvious that the BPS bound in the type IIB theory can not be saturated 
unless the parameter $h$ equals to zero.
Thus the $1/4$-susy BPS state in the ${\cal M}_{3+}$ case corresponds to 
a non-BPS string state, while the ${\cal N}_{3+}$ case gives a 
$1/2$-susy BPS string state.
\par

\vspace{0.5cm}
\noindent
[IV] ${\cal M}_{4+} = h K_L + E_R$ \\
This case and the ${\cal N}_{4+}=E_R$ case can be obtained by exchanging 
the subscripts $L \leftrightarrow R$ in [III].
All of results on possible BPS states in the ${\cal M}_{4+}$ (${\cal N}_{4+}$) 
case are the same as ones in the ${\cal M}_{3+}$ (${\cal N}_{3+}$) case, 
except that the signs of $Z_{\t010}$ and $Z_{910}$ are opposite to those in 
[III].
Hence, through the Weyl projection, we have $1/4$ and $3/4$-susy
states in [III] with the opposite signs of $P_{10}$ and $Y_{910}$,
while, through the IIB projection, we obtain string states in [III]
with the opposite sign of ${\tilde Z}^{ii}_9$.
\par 

\vspace{0.3cm}
The above ${\cal S}_2$, ${\cal N}_{3+}$ and ${\cal N}_{4+}$ cases were 
previously discussed in \cite{sh}.
In summary, we give the list of all possible BPS states in Table 3.
\par
\newpage
\begin{center} Table 3. \\
\begin{tabular}{|c|cccc|} \hline
BPS state & Susy & M-susy & IIA-susy & IIB-susy \\
${\cal S}_2$ = (2+2)-brane & $\hf$ & $\hf$ & $\hf$ & $\hf$ \\
${\cal N}_{1+} = E_L + E_R$ & $\hf$ & $\hf$ & $\hf$ & no-susy \\
${\cal M}_{3+}= E_L + h K_R$& ${1 \over 4}$ &
${1 \over 4}$ & ${1 \over 4}$ & no-susy \\
${\cal N}_{3+} = E_L$ & ${3 \over 4}$ & ${3 \over 4}$
& ${3 \over 4}$ & $\hf$ \\
${\cal M}_{4+}= h K_L + E_R$ & ${1 \over 4}$ &
${1 \over 4}$ & ${1 \over 4}$ & no-susy \\
${\cal N}_{4+} = E_R$ & ${3 \over 4}$ & ${3 \over 4}$
& ${3 \over 4}$ & $\hf$
\\ \hline 
\end{tabular} 
\end{center}	
\par 

\subsection{$so(10,2)$ case}%
Finally we show the classification of the full $so(10,2)$ case.
All types $\dlt$ in $so(10,2)$ can be divided into two cases;
[I] the $so(2,2)$ part in Sec.\,3.3 plus the remaining $so(8)$ 
part and [II] other cases which can not be decomposed as in [I].
\par

\vspace{0.2cm}
\noindent
[I] $so(2,2)$ part + $so(8)$ part \\
Since the group $SO(8)$ is compact, all types $\dlt$ in this case 
are given by the sum of types in $so(2,2)$ and the $so(8)$ type with
the canonical form as its representative,
\be
\dlt_{so(10,2)} = \dlt_{so(2,2)} + 
\sum_{k=3}^6 \dlt_0^+(i h_k, -i h_k) \ ,
\ee
where $dim.\dlt_{so(10,2)}=12$ and $ind.\dlt_{so(10,2)}=2$, as 
$dim.\dlt_0^+=2$ and $ind.\dlt_0^+=0$.
A representative for each type $\dlt_{so(10,2)}$ is therefore obtained by 
arranging a representative of each $\dlt_{so(2,2)}$ and four $2 \times 2$ 
anti-symmetric blocks $i h_k \sg_2$ $(k=3,4,5,6)$ on the diagonal part.
\par

\vspace{0.5cm}
\noindent
[I-1] $\displaystyle {\cal S}_2(h_1,h_2) + \sum_{k=3}^6 i h_k \sg_2$ \\
This case with ${\cal S}_2$ in Sec.\,3.3 corresponds to the (10+2)D canonical 
form and is naturally interpreted as the configuration of super (2+2)-branes 
extended in timelike $(\t0,0)$- and spacelike (1,2)-, (3,4)-, (5,6)-, (7,8)-, 
(9,10)-directions.
The calculation of the characteristic polynomial gives
\be
D(x) = \prod_{\pm}^{2^5} 
(h_1 \pm h_2 \pm h_3 \pm h_4 \pm h_5 \pm h_6 - x)^2 \ ,
\label{12cf}
\ee
where $h_1 = Z_{\t0 0} \geq 0$ and the above expression means the product 
of all possible combinations of the signs $\pm$. 
Since the total number of combinations equals to $2^5 = 32$, the RHS of 
(\ref{12cf}) becomes a polynomial of order 64 of the parameter $x$, as 
expected.
\par

It is easy to evaluate all possible fractions of survived 
supersymmetries in (\ref{12cf}).
At first, let parameters $h_i$ $(i=2,\cdots,6)$ be non-zero.
Then the unique minimal eigenvalue among 
$x= h_1 \pm h_2 \pm h_3 \pm h_4 \pm h_5 \pm h_6$ is 
\be
x_{min.} \equiv h_1 - |h_2| - |h_3| - |h_4| - |h_5| - |h_6| \ ,
\label{min}
\ee
and thus, from the positivity condition, we have the bound,
\be
h_1 \geq |h_2| + |h_3| + |h_4| + |h_5| + |h_6| \ .
\label{bnd}
\ee
A BPS state arises when the bound is saturated and, as each eigenvalue 
enters twice in the RHS of (\ref{12cf}), the fraction of surviving 
supersymmetry is $1/32$.
Next, we suppose that one of parameters, say $h_6$, is zero.
In this case, two eigenvalues $x_{mim.}$ and $x_{mim.}+2|h_6|$ become minimal 
and hence we have a BPS state with $1/16$-susy.
More generally, we obtain a $1/32$, $1/16$, $1/8$, $1/4$, $1/2$-susy 
BPS state if the number of zero $h_i$'s is $0,1,2,3,4$, respectively.
\par

BPS states in the $N=(1,0)$ superalgebra can be found by taking the 
Weyl projection of $D(x)$,
\be
D_W(x) =
\prod_{- : {\rm even}}^{2^4} 
(h_1 \pm h_2 \pm h_3 \pm h_4 \pm h_5 \pm h_6 - x)^2 \ ,
\label{12wcf}
\ee
where all combinations with even number of negative signs
are taken in the RHS.
Here, we should be careful for finding zero eigenvalues, 
since the minimal value $x_{min.}$ in (\ref{min}) may or may not
be contained in the RHS of (\ref{12wcf}). 
The complete classification of BPS states in this Weyl-projected case has 
been done in \cite{mmm}.
\par

If one of $h_i$ $(i=2,\cdots,6)$ is zero, $x_{min.}$ always belongs to 
the set of eigenvalues in (\ref{12wcf}) and other eigenvalues are obtained 
by adding $2 |h_{i_1}| + 2 |h_{i_2|}$ or $2 \sum_{q=1}^4 |h_{i_q}|$ to 
$x_{min.}$.
As in the case (\ref{12cf}), we can easily count the number of eigenvalues
equal to $x_{min.}$ and obtain a BPS state with $1/16$, $1/8$, $1/4$, 
$1/2$-susy, when the number of zero $h_i$'s is $1,2,3,4$, respectively.
On the other hand, let us suppose all $h_i$ to be non-zero. 
If $x_{min.}$ is contained in (\ref{12wcf}), then $x_{min.}$ is the 
unique minimal eigenvalue and its associated BPS state has the 
$1/16$-susy.
If $x_{min.}$ is not in (\ref{12wcf}), the minimal eigenvalue in (\ref{12wcf}) 
is given by adding some $+2|h_k|$ to $x_{min.}$, where the parameter $h_k$ 
is supposed to have a minimal absolute value among $h_i$'s.
If there are $q$ $h_i$'s with the minimal absolute value, the fraction of 
survived supersymmetry becomes $q/16$ $(q=1,\cdots,5)$ when the BPS bound
is saturated.
In summary, in the (10+2)D N=(1,0) superalgebra, there arise
$1/16$, $1/8$, $3/16$, $1/4$, $5/16$ and $1/2$-susy BPS states
from the second-rank central term.
\par

\vspace{0.5cm}
\noindent 
[I-2] $\displaystyle {\rm Other} \ so(2,2) \ {\rm cases} + 
\sum_{k=3}^6 i h_k \sg_2$\\
There also appear types $\dlt$ in $so(10,2)$ with ${\cal N}_{1+}$, 
${\cal M}_{3+}$, ${\cal N}_{3+}$, ${\cal M}_{4+}$ and ${\cal N}_{4+}$ 
in the $so(2,2)$ part.
Characteristic polynomials for the types become
\bea
&{\cal N}_{1+}:& D(x) = (D_W(x))^2 \ , \ 
D_W(x) = \prod_{\pm}^8 
(x(x-2) - (h_3 \pm h_4 \pm h_5 \pm h_6)^2)^2  \ , 
\label{ch1}\\
&{\cal M}_{3+}: & D(x) = D'_W(x) D_W(x) \ , 
\nonumber \\
& & D_W(x) = \prod_{-:{\rm odd}}^4 (x(x-2) - 
(h_3 \pm h_4 \pm h_5 \pm h_6)^2)^2\times 
\nonumber \\
& & \qquad \qquad \ 
\prod_{-:{\rm even}}^8 (x - 2 h \pm h_3 \pm h_4 \pm h_5 \pm h_6)^2 \ , 
\label{ch2}\\
& & D'_W(x) = \prod_{-:{\rm even}}^4 (x(x-2) - 
(h_3 \pm h_4 \pm h_5 \pm h_6)^2)^2 \times 
\nonumber \\
& & \qquad \qquad \
\prod_{-:{\rm odd}}^8 (x - 2 h \pm h_3 \pm h_4 \pm h_5 \pm h_6)^2 \ , 
\nonumber \\
&{\cal M}_{4+}: & D(x) = D'_W(x) D_W(x) \ , 
\nonumber \\ 
& &D_W(x) = D'_W(x) \ {\rm in} \ {\cal M}_{3+} \ , \quad  
D'_W(x) = D_W(x) \ {\rm in} \ {\cal M}_{3+} \ .
\label{ch3}
\eea
${\cal N}_{3+}$ and ${\cal N}_{4+}$ cases are obtained by setting
$h=0$ in the ${\cal M}_{3+}$ and ${\cal M}_{4+}$ cases,
respectively.
{}From (\ref{ch1}), (\ref{ch2}), (\ref{ch3}), it can be shown that
all $h_k$ $(k=3,4,5,6)$ in the $so(8)$ part have to be zero in all 
above cases to preserve the positivity condition.
Therefore, unlike the [I-1] case, no further fraction of survived 
supersymmetry can be obtained by adding the parameters $h_k$.
\par

\vspace{0.5cm}
\noindent
[II] Other $so(10,2)$ cases\\
Adding to the types in [I], there arise two types in the 
classification of $so(10,2)$ which contain indecomposable 
types with the dimension higher than four, that is, 
${}^{(\ep_D,\ep_C)}\dlt^-_2(ih,-ih)$ with $dim.\dlt = 6$ and 
$ind.\dlt = 2$, and ${}^{(+,\ep_C)}\dlt_4^+(0)$ with $dim.\dlt = 5$ 
and $ind.\dlt = 2$.
\par

\vspace{0.5cm}
\noindent
[II-1] $\displaystyle {}^{(\ep_D)}{\cal F}_1 = 
{}^{(\ep_D,+)}\dlt_2^- (i h, -i h) + \sum_{k=2}^4 \dlt_0^+ (i h_k, -i h_k)$\\
The type $\dlt_2^- (i h, -i h)$ under $O(10,2)$ splits into four distinct
parts under the action of $SO(10,2)_0$.
For the $C$- and $D$-splittings, we use the following $6\times6$ matrices,
\be
C =
\left( \begin{array}{ccc}
\sg_3 \ & 0 \ & 0 \ \\ 0 \ & \sg_3 \ & 0 \ \\ 0 \ & 0 \ & I_2 \ 
\end{array} \right) \ , \quad
D =
\left( \begin{array}{ccc}
I_2 \ & 0 \ & 0 \ \\ 0 \ & I_2 \ & 0 \ \\ 0 \ & 0 \ & \sg_3 \ 
\end{array} \right) \ ,
\label{6cd}
\ee
which act on a representative of $\dlt_2^- (i h, -i h)$ given in (\ref{a13})
and yield the representative ${}^{(\ep_D,\ep_C)}Z$ of the type
${}^{(\ep_D,\ep_C)}\dlt_2^- (i h, -i h)$ with respect to the (4+2)D flat 
metric $\eta = (-,-,+,+,+,+)$,
\be
{}^{(\ep_D,\ep_C)}Z =
\left( \begin{array}{cccccc}
0 & -h & \hsf & 0 & 0 & 0 \\ h & 0 & 0 & \hsf & 0 & 0 \\
\hsf & 0 & 0 & -h & \hsf & 0 \\ 
0 & \hsf & h & 0 & 0 & {\ep_D \over {\sqrt 2}}\\
0 & 0 & -\hsf & 0 & 0 & -\ep_D h \\ 0 & 0 & 0 & - {\ep_D \over {\sqrt 2}} & 
\ep_D h & 0 \end{array} \right) \ ,
\ee
where $\ep_C = +1 \, (-1)$ corresponds to the range $h > 0 \, (< 0)$, showing
the $\ep_C=+1$ case to be compatible with the positivity condition.
Arranging the matrix ${}^{(\ep_D,+)}Z$ and $i h_k \sg_2$ $(k=2,3,4)$ on the 
diagonal part, we obtain a representative of the type
${}^{(\ep_D)}{\cal F}_1$ with respect to our (10+2)D flat metric $\eta_{AB}$.
The characteristic polynomial for the type is 
\bea
&& D(x) = D'_W(x)D_W(x) \ , \qquad  
D_W(x) = D_{\rho=-1}(x) \ , \quad D'_W(x) = D_{\rho=1}(x) \ , 
\nonumber \\
&& D_{\rho} = \prod_{- :{\rm even}}^4 
(x - 3 h + \rho \ep_D (\pm h_2 \pm h_3 \pm h_4))^2 \times 
\nonumber \\
&&((x - h - \rho \ep_D (\pm h_2 \pm h_3 \pm h_4))^2
(x + h + \rho \ep_D (\pm h_2 \pm h_3 \pm h_4)) - 4 x)^2 \ .
\label{ii-1}
\eea
It can be shown that, for any values of $h$, $h_2$, $h_3$ and $h_4$, there 
appears at least one negative eigenvalue from the cubic polynomial part 
of the parameter $x$ in the last line of (\ref{ii-1}).
Therefore the type ${}^{(\ep_D)}{\cal F}_1$ can not be used to define 
anti-commutators among supercharges.
\par

\vspace{0.5cm}
\noindent
[II-2] $\displaystyle {\cal F}_2 = {}^{(+,+)}(\dlt_4^+(0) + \dlt_0^+(0)) + 
\sum_{k=2}^4 \dlt _0^+(i h_k, -i h_k)$\\
The type $\dlt_4^+(0) + \dlt_0^+(0)$ under $O(10,2)$ splits into two parts 
${}^{(+,\ep_C)}(\dlt_4^+(0) + \dlt_0^+(0))$ under $SO(10,2)_0$.
A representative in $\dlt_4^+(0)$ is constructed in (\ref{a33}), which
gives, after the $C$-slipping generated by the $C$ matrix in (\ref{6cd}),
the representative ${}^{(\ep_C)}Z$ in 
${}^{(+,\ep_C)}(\dlt_4^+(0) + \dlt_0^+(0))$ with respect to the (4+2)D 
metric $\eta = (-,-,+,+,+,+)$, 
\be
{}^{(\ep_C)}Z = {\ep_C \over 2}
\left( \begin{array}{cccccc}
0 \ & -1 \ & 0 \ & 1 \ & 0 \ & 0 \\
1 \ & 0 \ & 1 \ & 0 \ & {\sqrt 2} \ & 0 \\
0 \ & 1 \ & 0 \ & -1 \ & 0 \ & 0 \\ 
1 \ & 0 \ & 1 \ & 0 \ & -{\sqrt 2} \ & 0 \\
0 \ & {\sqrt 2} \ & 0 \ & {\sqrt 2} \ & 0 \ & 0 \\ 
0 \ & 0 \ & 0 \ & 0 \ & 0 \ & 0 
\end{array} \right) \ ,
\ee
showing $\ep_C = -1$ to be forbidden by the positivity condition.
We obtain a representative of the type ${\cal F}_2$ with respect to 
the (10+2)D flat metric $\eta_{AB}$ by arranging the matrix ${}^{(+)}Z$ and 
$i h_k \sg_2$ $(k=2,3,4)$ on the diagonal block.
The characteristic polynomial in this ${\cal F}_2$ case becomes
\bea
&& D(x) = (D_W(x))^2 \ , \nonumber \\
&& D_W(x) = \prod_{-: {\rm even}}^4
((x^2 - 2 - (\pm h_2 \pm h_3 \pm h_4)^2) (x (x - 2) 
- (\pm h_2 \pm h_3 \pm h_4)^2) \nonumber \\
&& \qquad \qquad \qquad \ 
- 2 (\pm h_2 \pm h_3 \pm h_4)^2)^2 \ .
\label{ii-2}
\eea
As in the [II-1] case, for any values of $h_2$, $h_3$ and $h_4$, we have 
at least one negative eigenvalue as a solution of the fourth order equation 
of the parameter $x$ in the RHS of (\ref{ii-2}).
Hence the type ${\cal F}_2$ can not be used for the definition of 
anti-commutators among supercharges.
\par

\vspace{0.2cm}
With the above two cases, the classification of the $so(10,2)$ algebra under 
$SO(10,2)_0$ has been completed and all possible BPS states arising 
from the central term $Z_{AB}$ have been exhausted.
\par

\section{Conclusion}%
In summary, we have studied the $N=(1,1)$ non-Poincar\'e superalgebra 
in (10+2)D with tensorial central terms $Z^{(k)}$ $(k=2,3,6,7,10,11)$.
The Weyl-projected form of the algebra, i.e., the $N=(1,0)$ superalgebra, has
been shown to be reduced to the (10+1)D M-algebra by a timelike dimensional 
reduction, just as the M-algebra is reduced to the (9+1)D type IIA algebra
by a spacelike dimensional reduction.
Another $SO(9,1)$ covariant projection of the $N=(1,1)$ algebra with 
dimensional reduction to (9+1)D has also been demonstrated to yield the 
type IIB algebra.
\par

{}From the complete classification of the second-rank central term, we have 
exhausted all possible BPS states arising from this term.
A $1/2$-susy BPS state is associated with the semisimple ${\cal S}_2$ type 
in $so(2,2)$ and is naturally interpreted as a super (2+2)-brane, which is
dimensionally reduced to a membrane in (10+1)D and type IIA and IIB 
fundamental strings in (9+1)D.
Further fractions less than $1/2$ of surviving supersymmetries have been
obtained from the extension of the ${\cal S}_2$ type to the canonical type 
in $so(10,2)$.
There has arisen another $1/2$-susy BPS state from the nilpotent 
${\cal N}_{1+}$ type in $so(2,2)$, which would be observed as an M9 brane 
in M-theory.
The nilpotent type has been caused due to the non-compact property of the 
group $SO(10,2)$.
Thus the further investigation of the type might give some insight or 
constraint on the number of time directions of spacetime.
We also have obtained the fraction $1/4$ of surviving supersymmetries in 
${\cal M}_{3+}$ and ${\cal M}_{4+}$ types and the fraction $3/4$ in 
${\cal N}_{3+}$ and ${\cal N}_{4+}$ types.
As discussed in Sec.\,3.3.2 [III], if BPS states with these fractions of 
supersymmetry can possibly exist in M-theory, they would be realized
as some composite states of M2 and M9 branes.
In recent paper \cite{gh}, it has been shown that a configuration of
an M2 brane intersecting two M5 branes also preserves $1/4$ or $3/4$
of supersymmetry when the product of all three brane charges is 
negative.
It remains to investigate whether there exist (10+1)D supergravity
solutions corresponding to those $1/4$ and $3/4$-susy BPS states, 
or not.
\par

Besides the second-rank central term, the consideration of other central
terms leads us to other various brane configurations inherent in the 
(10+2)D superalgebra.
Among them, the sixth-rank central term $Z^{(6)}$ represents a super 
(6+2)-brane, which is dimensionally reduced to an M5 brane in M-theory
and an NS5 brane in the type IIB theory.
It has been discussed that the worldvolume theory of the (2+2)-brane moving
in (10+2)D spacetime is given by (2+2)D self-dual Yang-Mills and 
gravitational theories \cite{km}.
It is intriguing to study whether the (6+2)-brane is described by some 
(6+2)D integrable systems like the self-dual theories.
A candidate would be a (6+2)D version of the 8D self-dual Yang-Mills theory 
\cite{cdfn} and its gravitational analogue \cite{ao}.
\par

The method used in this paper to classify the central term $Z^{(2)}_{AB}$ 
under the $SO(10,2)$ rotation is also applicable to any second-rank central 
term in any dimension and signature of spacetime.
An example is the $N=(2,0)$ chiral superalgebra in (10+2)D, which is a
non-Poincar\'e type algebra containing central terms $Z^{(2)}$, $Z$, $Z^{(4)}$ 
and self-dual ${}^{(+)}Z^{(6)}$ \cite{rss},
\bea
\{Q^{\al}_i, Q^{\beta}_j\} &=& 
(\tau^k)_{ij}((\Gm^{AB}C^{-1})^{\al\beta}Z^{(2)}_{k\,AB}+
(\Gm^{A_1 \cdots A_6}C^{-1})^{\al\beta}{}^{(+)}Z^{(6)}_{k\,A_1\cdots A_6})
\nonumber \\
& &+ \ep_{ij} ((C^{-1})^{\al\beta} Z + (\Gm^{A_1\cdots A_4}C^{-1})^{\al\beta}
Z^{(4)}_{A_1 \cdots A_4}) \ ,
\label{cqq}
\eea
where $Q^{\al}_i$ $(i=1,2)$ are Majorana-Weyl supercharges with the same 
chirality and $\tau^k = (\sg_3, \sg_1, I_2)$.
As noted in the introduction, like the $N=(1,1)$ case, the $N=(2,0)$ algebra
also reduces to M-, type IIA and IIB algebras by dimensional reduction.
It is interesting to find what types of BPS states exist in this chiral 
algebra and to clarify the connection of them with various BPS states in 
membrane and string theories.
\par

\newpage
\section*{Acknowledgements}
I thank N. Kawanaka for useful suggestions.
I am grateful to H. Kihara and D.B. Fairlie for discussions and 
helpful comments.
I also thank the department of Mathematical Sciences in University
of Durham and the department of Mathematics in King's College, London
for their hospitality.

\vspace{1.0cm}
\section*{Appendix}%
According to the description of indecomposable types in Sec.\,3.2.2, we 
construct explicit forms of representatives for the types $\dlt_m$ relevant 
to our $o(10,2)$ case.
Let $(Z,V) \in \dlt_m$ and $Z$ be decomposed as the sum of semisimple $S$ 
and nilpotent $N$ elements, $Z= S+N$, with $N^{m+1}=0$.
The vector space $V$ can be written as $V = F+NF+\cdots+N^mF$, where
$F$ is the subspace of $V$ spanned by eigenvectors of $S$.
{}From (\ref{so}), the metric $\tau_m(u,v) = \tau(u, N^mv)$ for $u,v \in F$ 
is symmetric for $m=$ even and anti-symmetric for $m=$ odd.
For later use, we define the symmetric, nondegenerate and real bilinear form
$\tht_m$ as 
\bea
\tht_m(u,v) &=& \tau_m(u,v) \qquad \ \ {\rm if} \ m={\rm even}, 
\nonumber \\
&=& \tau_m(u,Sv) \qquad {\rm if} \ m={\rm odd} \ {\rm and} \ S \not=0.
\eea
\par

\subsection*{A-1: $\dlt^\ep_m(ih,-ih)$}%
We can choose a pair of vectors $\{e,f\}$ spanning the subspace 
$F$ of $V$ such that $Se=-hf$ and $Sf=he$ ($h \in R$).
{}From $\tht_m(Su,v)+\tht_m(u,Sv)=0$, we have $\tht_m(e,f)=0$ while
$\tht_m(e,e)=\tht_m(f,f)$.
We have two choices of the sign of $\tht_m(e,e)$, which is labeled by
the superscript $\ep=\pm1$ in the type.
\par

\subsubsection*{A-1-1: m=0 \ \ $dim.\dlt=2$}
Since $N^{0+1}=0$, this type contains only a semisimple element $Z=S$.
The form of $S$ on $V= F= \{e,f\}$ with respect to (w.r.t.) the metric $\tau$ 
becomes
\be
Z= S = \left( 
\begin{array}{cc}0 \ & h  \ \\ -h \ & 0 \ 
\end{array} \right) \ , \qquad {\rm w.r.t.} \ \ \ 
\tau = \left( 
\begin{array}{cc} \ep \ & 0  \ \\ 0 \ & \ep \ 
\end{array} \right) \ .
\ee
The form of the metric shows that the type $\dlt^+$ has the index 0, while 
$\dlt^-$ the index 2.
\par

\subsubsection*{A-1-2: m=1 \ \ $dim.\dlt=4$, \ $ind.\dlt=2$}
Adding to a semisimple element $S$, we have a nilpotent $N$ with $N^2=0$. 
A representative in this case is given as 
\be
Z=S+N \ , \quad 
S = h \left( 
\begin{array}{cc}-\ve_2 \ & 0  \ \\ 0 \ & -\ve_2 \ 
\end{array} \right) \ , \quad \ \
N = \left( 
\begin{array}{cc}0 \ & 0 \ \\ I_2 \ & 0 \ 
\end{array} \right) \ ,
\ee
w.r.t. the metric
\be
\tau = {\ep \over h} \left( 
\begin{array}{cc} 0 \ & \ve_2  \ \\ -\ve_2 \ & 0 \ 
\end{array} \right) \ .
\ee
We perform a transformation of $V$ onto $V'$ by an isomorphism $\phi$ 
noted in Sec.\,3.2.2 to make the form of the metric $\tau$ to be 
$\tau' = \phi \tau {\tilde \phi} = (-,-,+,+)$.
With a realization $O_{\phi}$ of $\phi$, we have
\be
Z'= {\tilde O}_{\phi} Z O_{\phi} = \hf 
\left( 
\begin{array}{cc} -(2h+1)\ve_2 \ & \ve_2  \ \\ -\ve_2 \ & -(2h-1)\ve_2 \ 
\end{array} \right) \ , \ \ 
O_{\phi} = \sqrt{{h \over 2}} \left( 
\begin{array}{cc} I_2 \ & -I_2  \ \\ \ve_2 \ & \ve_2 \ 
\end{array} \right) \ ,
\ee
where $h>0$ ($h<0$) corresponds to $\ep=+1\,(-1)$.
\par

\subsubsection*{A-1-3: m=2 \ $\ep=-1$, \ \ $dim.\dlt=6$, \ $ind.\dlt=2$}
Since the $\ep=+1$ case has $ind.\dlt=4$, only the $\ep=-1$ case is relevant 
to our $o(10,2)$ algebra.
The form of a representative in the $\ep= -1$ case becomes
\be
Z=S+N \ , \quad 
S = h \left( 
\begin{array}{ccc}-\ve_2 \ & 0 \ & 0 \ \\ 0 \ & -\ve_2 \ & 0 \ \\
0 & 0 & -\ve_2 \ \end{array} \right) \ , \quad \ 
N = \left( 
\begin{array}{ccc}0 \ & 0 \ & 0 \ \\ I_2 \ & 0 \ & 0 \ \\
0 \ & I_2 \ & 0 \ \end{array} \right) \ ,
\ee
w.r.t. the metric 
\be
\tau = \left( 
\begin{array}{ccc} 0 \ & 0 \ & -I_2 \ \\ 0 \ & I_2 \ & 0 \ \\
-I_2 \ & 0 \ & 0 \ \end{array} \right) \ .
\ee
Performing the $\phi$-transformation $O_{\phi}$, we have a new representative
\be
Z' = {\tilde O}_{\phi} Z O_{\phi} = \hsf \left( 
\begin{array}{ccc} 
-{\sqrt 2}h\ve_2 \ & I_2 \ & 0 \ \\ I_2 \ & - {\sqrt 2}h\ve_2 \ & I_2 \ \\
0 \ & - I_2 \ & -{\sqrt 2}h\ve_2 \ \ \end{array} \right) \, , \ 
O_{\phi} = \hsf \left( 
\begin{array}{ccc} I_2 \ & 0 \ & I_2 \ \\ 0 \ & {\sqrt 2}I_2 \ & 0 \ \\
I_2 & 0 & -I_2 \ \end{array} \right) \ ,
\label{a13}
\ee
w.r.t. $\tau' = (-,-,+,+,+,+)$.
\par

\subsection*{A-2: $\dlt_m(h,-h)$}%
As eigenvalues of $S$ are real, we can take eigenvectors $\{e,f\}$ such that
$Se=he$ and $Sf=-hf$.
We can derive $\tht_m(e,e)=\tht_m(f,f)=0$ and, without restriction, take 
$\{e,f\}$ to satisfy $\tht_m(e,f)=1$.
\par

\subsubsection*{A-2-1: m=0 \ \ $dim.\dlt=2$, \ $ind.\dlt=1$}
This case contains only a semisimple element $S$ w.r.t. the metric $\tau$,
\be
S = \left( 
\begin{array}{cc}h \ & 0  \ \\ 0 \ & -h \ 
\end{array} \right) \ , \qquad {\rm w.r.t} \ \ \ 
\tau = \left( 
\begin{array}{cc} 0 \ & 1  \ \\ 1 \ & 0 \ 
\end{array} \right) \ .
\ee
By the $\phi$-transformation $O_{\phi}$, we have
\be
S' = {\tilde O}_{\phi} S O_{\phi} = \left( 
\begin{array}{cc} 0 \ & h  \ \\ h \ & 0 \ 
\end{array} \right) \ , \quad
O_{\phi} = \hsf \left( 
\begin{array}{cc} 1 \ & 1  \ \\ -1 \ & 1 \ 
\end{array} \right) \ , 
\ee
w.r.t. $\tau' = (-,+)$.
\par

\subsubsection*{A-2-2: m=1 \ \ $dim.\dlt=4$, \ $ind.\dlt=2$}
This case contains a nilpotent $N$ with $N^2=0$, adding to a semisimple $S$,
\be
Z=S+N \ , \quad 
S = h \left( 
\begin{array}{cc}\sg_3 \ & 0  \ \\ 0 \ & \sg_3 \ 
\end{array} \right) \ , \quad
N = \left( 
\begin{array}{cc}0 \ & 0 \ \\ I_2 \ & 0 \ 
\end{array} \right) \ ,
\label{a22z}
\ee
w.r.t. the metric
\be
\tau = \left( 
\begin{array}{cc} 0 \ & -\ve_2  \ \\ \ve_2 \ & 0 \ 
\end{array} \right) \ .
\label{a22g}
\ee
By the $\phi$-transformation $O_{\phi}$, we obtain another representative  
\be
Z' = {\tilde O}_{\phi} Z O_{\phi} = \hf \left( 
\begin{array}{cc} \ve_2 \ & 2h\sg_3+\ve_2  \ \\
2h\sg_3-\ve_2  \ & -\ve_2 \ 
\end{array} \right) \, , \ 
O_{\phi} = \hsf \left( 
\begin{array}{cc} I_2 \ & I_2  \ \\ -\ve_2 \ & \ve_2 \ 
\end{array} \right) \, , 
\label{a22f}
\ee
w.r.t. $\tau'= (-,-,+,+)$.
\par

\subsection*{A-3: $\dlt^{\ep}_m(0)$}%
In this case, there appears only a nilpotent $N$ with $N^{m+1}=0$.
Let $e$ be a vector in $V$, then we have $V= e+Ne+\cdots+N^me$.
We can take $\tht_m(e,e)=\ep$ without any restriction, which determines
the form of the metric completely.
\par

\subsubsection*{A-3-1: m=0 \ \ $dim.\dlt=1$}
This case contains the trivial element $Z=(0)$ w.r.t. the 
metric $\tau=\ep$.
\par

\subsubsection*{A-3-2: m=2 \ \ $dim.\dlt=3$}
In this case, we have a $3\times3$ nilpotent matrix
\be
Z = N = \left( 
\begin{array}{ccc}0 \ & 0  \ & 0 \ \\ 1 \ & 0 \ & 0 \ \\
0 \ & 1 \ & 0 \
\end{array} \right) \ , \qquad {\rm w.r.t} \ \ \ 
\tau = \left( 
\begin{array}{ccc} 0 \ & 0 \ & \ep \ \\ 0 \ & -\ep \ & 0 \\
\ep \ & 0 \ & 0 \
\end{array} \right) \ .
\ee
In the $\ep=+1$ case, we obtain a representative in $\dlt^+_2(0)$ w.r.t. 
$\tau'=(-,-,+)$ by the $\phi$-transformation $O_{\phi}$,
\be
N' = {\tilde O}_{\phi} N O_{\phi} = \hsf
\left( 
\begin{array}{ccc} 0 \ & -1 \ & 0 \ \\ 1 \ & 0 \ & 1 \\
0 \ & 1 \ & 0 \
\end{array} \right) \ , \quad
O_{\phi} = \hsf \left( 
\begin{array}{ccc} 1 \ & 0 \ & 1 \ \\ 0 \ & {\sqrt 2} \ & 0 \ \\
-1 \ & 0 \ & 1 \ \end{array} \right) \ .
\ee
On the other hand, in the $\ep=-1$ case, by another $\phi$-transformation 
$O'_{\phi}$, we have a representative in $\dlt^-_2(0)$,
\be
N' = {\tilde O}'_{\phi} N O'_{\phi} = \hsf
\left( 
\begin{array}{ccc} 0 \ & 1 \ & 0 \ \\ 1 \ & 0 \ & 1 \\
0 \ & -1 \ & 0 \
\end{array} \right) \ , \quad
O'_{\phi} = \hsf \left( 
\begin{array}{ccc} 1 \ & 0 \ & 1 \ \\ 0 \ & {\sqrt 2} \ & 0 \ \\
1 \ & 0 \ & -1 \ \end{array} \right) \ ,
\ee
w.r.t. $\tau'= (-,+,+)$.
\par

\subsubsection*{A-3-3: m=4 \ $\ep=+1$, \ \ $dim.\dlt=5$, \ $ind.\dlt=2$}
The case with $\ep=-1$ is forbidden since it has $ind.\dlt=3$.
A representative for the $\ep=+1$ case is given as
\be
Z = N = \left( 
\begin{array}{ccccc} 0 \ & 0 \ & 0 \ & 0 \ & 0 \ \\ 
1 \ & 0 \ & 0 \ & 0 \ & 0 \ \\ 0 \ & 1 \ & 0 \ & 0 \ & 0 \ \\
0 \ & 0 \ & 1 \ & 0 \ & 0 \ \\ 0 \ & 0 \ & 0 \ & 1 \ & 0 \ 
\end{array} \right) \ , \qquad {\rm w.r.t} \ \ \ 
\tau = \left( 
\begin{array}{ccccc} 0 \ & 0 \ & 0 \ & 0 \ & 1 \ \\ 
0 \ & 0 \ & 0 \ & -1 \ & 0 \ \\0 \ & 0 \ & 1 \ & 0 \ & 0 \ \\
0 \ & -1 \ & 0 \ & 0 \ & 0 \\1 \ & 0 \ & 0 \ & 0 \ & 0 \
\end{array} \right) \ .
\ee
Performing the $\phi$-transformation $O_{\phi}$, we obtain another 
representative in $\dlt^+_4(0)$, 
\be
N' = {\tilde O}_{\phi} N O_{\phi} = \hf
\left( 
\begin{array}{ccccc} 0 \ & -1 \ & 0 \ & 0 \ & 1 \ \\ 
1 \ & 0 \ & {\sqrt 2} \ & 1 \ & 0 \ \\ 
0 \ & {\sqrt 2} \ & 0 \ & 0 \ & {\sqrt 2} \ \\
0 \ & 1 \ & 0 \ & 0 \ &-1 \ \\
1 \ & 0 \ & -{\sqrt 2} \ & 1 \ & 0
\end{array} \right) \ , \quad
O_{\phi} = \hsf \left( 
\begin{array}{ccccc} 1 \ & 0 \ & 0 \ & 1 \ & 0 \\ 
0 \ & 1 \ & 0 \ & 0 \ & 1 \\0 \ & 0 \ & {\sqrt 2} \ & 0 \ & 0 \ \\
0 \ & 1 \ & 0 \ & 0 \ &-1 \ \\ -1 \ & 0 \ & 0 \ & 1 \ & 0 
\end{array} \right) \ ,
\label{a33}
\ee
w.r.t. $\tau'=(-,-,+,+,+)$.

\par

\subsection*{A-4: $\dlt_1(0,0)$ \ \ $dim.\dlt=4$, \ $ind.\dlt=2$}%
This type gives a nilpotent $Z=N$ with $N^2=0$.
Let us take two vectors $\{e,f\}$ in $V$ as eigenvectors of $S$, then the 
vector space $V$ can be spanned by $\{e,f,Ne,Nf\}$.
We can derive $\tht_1(e,e)=\tht_1(f,f)=0$ and take $\tht_1(e,f)$ equal to $+1$.
The representative in $\dlt_1(0,0)$ and the metric $\tau$ derived from 
the setup are the same as those in (\ref{a22z}), (\ref{a22g}).
Using the $\phi$-transformation $O_{\phi}$ in (\ref{a22f}), we obtain another
representative w.r.t. the metric $\tau'=(-,-,+,+)$,
\be
N' = {\tilde O}_{\phi} N O_{\phi} = \hf \left( 
\begin{array}{cc} \ve_2 \ &\ve_2 \ \\ -\ve_2 \ & -\ve_2 \ 
\end{array} \right) \ .
\label{a4f}
\ee
\par

\subsection*{A-5: $\dlt_0(\zt,-\zt,{\bar \zt},-{\bar \zt})$ \ \ $dim.\dlt=4$, 
\ $ind.\dlt=2$}%
We have a semisimple $Z=S$ with eigenvalues
$\{\zt,-\zt,{\bar \zt},-{\bar \zt}\}$.
Let $\zt=h_1+ih_2$ and $\{e_{\pm},f_{\pm}\}$ be vectors in $V$ satisfying
\be
Se_{\pm} = \pm(h_1e_{\pm} - h_2f_{\pm}) \ , \quad
Sf_{\pm} = \pm(h_1f_{\pm} + h_2e_{\pm}) \ .
\ee
The values of $\tht_0$ for $\{e_{\pm},f_{\pm}\}$ are given as
\be
\tht_0(e_{\pm},e_{\pm}) =\tht_0(f_{\pm},f_{\pm}) = 0 \, , \ 
\tht_0(e_{\pm},f_{\pm}) =\tht_0(e_{\pm},f_{\mp}) = 0 \, , \ 
\tht_0(e_{\pm},e_{\mp}) = -\tht_0(f_{\pm},f_{\mp}) = 1 \, .
\ee
Explicit forms of $S$ and the metric $\tau$ w.r.t. $\{e_{\pm},f_{\pm}\}$ are
\be
S = \left( 
\begin{array}{cc}h_1\sg_3 \ & h_2\sg_3  \ \\ -h_2\sg_3 \ & h_1\sg_3 \ 
\end{array} \right) \ , \qquad {\rm w.r.t} \ \ \ 
\tau = \left( 
\begin{array}{cc} \sg_1 \ & 0  \ \\ 0 \ & -\sg_1 \ 
\end{array} \right) \ .
\ee
By the $\phi$-transformation $O_{\phi}$, we have another representative 
\be
S'= {\tilde O}_{\phi} S O_{\phi} = 
\left( 
\begin{array}{cc}-h_2\ve_2 \ & h_1I_2  \ \\ h_1I_2 \ & -h_2\ve_2 \ 
\end{array} \right) \ , \quad
O_{\phi} = \hsf \left( 
\begin{array}{cccc} 1 \ & 0 \ & 1 \ & 0 \ \\ 
-1 \ & 0 \ & 1 \ & 0 \ \\ 0 \ & 1 \ & 0 \ & 1 \ \\
0 \ & 1 \ & 0 \ &-1 \ \end{array} \right) \ ,
\ee
w.r.t. $\tau'= (-,-,+,+)$.

\vspace{0.5cm}
Having obtained the above representatives in indecomposable types, we arrange
them on the diagonal part of $12\times12$ matrix in order to obtain $o(10,2)$ 
representatives with $ind.\dlt=2$.
In general, we have to transform the obtained representatives into the other 
forms with respect to our (10+2)D metric $\eta=(-,-,+,\cdots,+)$.
Here we give an example in the $so(2,2)$ case in Sec.\,3.3.1.
Let us take the sum $\dlt = \dlt_0(h_1,-h_1)+\dlt_0(h_2,-h_2)$, which has 
$dim.\dlt=4$ and $ind.\dlt=2$.
{}From Sec.\,A-2-1, we have a representative $Z$ for the type $\dlt$ w.r.t. 
the metric $\tau=(-,+,-,+)$,
\be
Z = \left( 
\begin{array}{cc}h_1\sg_1 \ & 0  \ \\ 0 \ & h_2\sg_1 \ 
\end{array} \right) \ .
\ee
Then, by the $\phi$-transformation $O_{\phi}$, we have another representative 
in $\dlt$ w.r.t. our (2+2)D metric $\tau'= (-,-,+,+)$,
\be
Z'= {\tilde O}_{\phi} Z O_{\phi} = 
\left( 
\begin{array}{cccc} 0 \ & 0 \ & h_1 \ & 0 \\ 
0 \ & 0 \ & 0 \ & h_2 \\ h_1 \ & 0 \ & 0 \ & 0 \ \\
0 \ & h_2 \ & 0 \ & 0 \ \end{array} \right) \ , \quad \ \ 
O_{\phi} = \left( 
\begin{array}{cccc} 1 \ & 0 \ & 0 \ & 0 \ \\ 
0 \ & 0 \ & 1 \ & 0 \ \\ 0 \ & 1 \ & 0 \ & 0 \ \\
0 \ & 0 \ & 0 \ & 1 \ \end{array} \right) \ ,
\ee
which corresponds to the ${\cal S}_1$ case in Sec.\,3.3.1.
\par

\vspace{0.2cm}
\newpage

\end{document}